\documentclass[twocolumn,aps,prc,superscriptaddress,showpacs,nobibnotes]{revtex4}
\usepackage{epsfig,dcolumn,bm}

\newcommand{\sqd}{\sqrt{2}} 
\newcommand{\sqs}{\sqrt{6}} 

\begin{document}

\title{Radiative decay of the dynamically generated open and hidden charm scalar meson
resonances $D^*_{s0} (2317)$ and $X(3700)$ }

\author{D. Gamermann}{\thanks{E-mail: daniel.gamermann@ific.uv.es}
\affiliation{Research Center for Nuclear Physics (RNCP),Osaka University, Ibaraki
567-0047,Japan }
\affiliation{Departamento de F\'isica Te\'orica and IFIC, Centro Mixto
Universidad de Valencia-CSIC,\\ Institutos de Investigaci\'on de
Paterna, Aptdo. 22085, 46071, Valencia, Spain} 
\author{L. R. Dai}{\thanks{E-mail: dailr@lnnu.edu.cn}
\affiliation{Research Center for Nuclear
Physics (RNCP),Osaka University, Ibaraki 567-0047,Japan }
\affiliation{Department of Physics, Liaoning Normal University,
 Dalian, 116029, China}
\author{E. Oset}{\thanks{E-mail: oset@ific.uv.es}
\affiliation{Research
Center for Nuclear Physics (RNCP),Osaka University, Ibaraki
567-0047,Japan }
\affiliation{Departamento de F\'isica Te\'orica and IFIC, Centro Mixto
Universidad de Valencia-CSIC,\\ Institutos de Investigaci\'on de
Paterna, Aptdo. 22085, 46071, Valencia, Spain}

\begin{abstract}
We present the formalism for the decay of dynamically generated
scalar mesons with open- or hidden-charm and give results for the decay of
$D^*_{s0} (2317)$ to $\gamma D_s^*$ plus that of a hidden charm scalar meson state predicted
by the theory around 3700 MeV decaying into $\gamma J/\psi$.
\end{abstract}



\maketitle

\section{Introduction}

The discovery of charmed scalar meson resonances at Babar, Belle and CLEO
\cite{Aubert:2003fg,Krokovny:2003zq,Abe:2003zm,cleo} has stimulated a
fruitful line of research, suggesting that their structure is much
richer than what one might guess assuming the $q\bar{q}$ picture,
which has proved quite successful in other areas
\cite{Godfrey:1985xj}. Some authors have suggested a 
$qq\bar{q}\bar{q}$ structure \cite{Terasaki:2003qa,Wang:2006uba} or
mixing  between the usual $q\bar{q}$ structure and four quark
\cite{Vijande:2006hj}.  Also there have been suggestions that these
states might be molecular states of the pseudoscalar mesons
\cite{Szczepaniak:2003vy, Barnes:2003dj, Sassen:2005ej,
Weinstein:1990gu,beveren}. Similar to this latter works, but with subtle
differences that we shall discuss later on, are the pictures where these
states appear as dynamically generated in the context of
unitarized chiral perturbation theory \cite{Kolomeitsev:2003ac,
Hofmann:2003je,Guo:2006fu,daniel}. The works in
\cite{Kolomeitsev:2003ac,Hofmann:2003je,Guo:2006fu}
rely upon Lagrangians based on heavy quark symmetry, while the
one of \cite{daniel} starts from an extension of the $SU(3)$ chiral
Lagrangian to $SU(4)$ which is largely broken due to the implicit vector
meson exchange characterizing the Weinberg-Tomozawa term of the chiral
Lagrangian \cite{Gasser:1984gg}. Because of this, the terms of the Lagrangian used in
\cite{daniel} are suppressed by the ratio of the light to heavy mass
squared of the vector mesons. 
$SU(4)$ Lagrangians with covariant derivatives adapted to include weak interactions
are used in \cite{noguera}.
The Lagrangian in \cite{daniel} contains the Lagrangian used in
\cite{Kolomeitsev:2003ac,Guo:2006fu},
which is suited for the study of open-charm resonances like the
$D^*_{s0} (2317)$, but it also contains other terms which allow to study the
hidden-charm states. The comparison of the results using this
Lagrangian with another one using a chiral symmetry breaking
extension to $SU(N)$ of the SU(3) results \cite {dani34} allows to
have an idea of the uncertainties in the results. The stability of
the $D^*_{s0} (2317)$ was confirmed, while the $X(3700)$ state could
sometimes become a cusp instead of a bound state, but experimentally it
would lead to a bump in the mass distributions
 in any case. Hence one is confident that
the new state should also be found.

The radiative decay of resonances has been usually suggested as a test
for their nature \cite {wzg07,hanhart}. In the present case, we are interested
in the decay $D_{s0}^{*}(2317)\rightarrow D_{s}^{*}\gamma$ and $X
(3700) \rightarrow \gamma J/\psi$. The first of these decays has
been evaluated assuming varied structures for the resonance,
within quark models, vector meson dominance, light cone QCD sum rules, etc
\cite{Godfrey:2003kg,Colangelo:2003vg,Bardeen:2003kt,
:2003dpa,Ishida:2003gu,Azimov:2004xk,Colangelo:2005hv,Close:2005se,Liu:2006jx,Wang:2006zw}
and more recently from the point of view of the $D_{s0}^{*}(2317)$ as a
molecular state \cite {prd76}. The $X(3700)$ has also been predicted
in \cite{Zhang:2006ix} as a molecular state assuming a reasonable
interaction Lagrangian between the $D$ and $\bar{D}$ states. The radiative
decay of this predicted state is reported here for the first time.

\section{Formulation for the Radiative Decay}

In a picture of the scalar mesons as dynamically generated one
needs to couple the photon to the meson components of the coupled
channels. Here the reactions studied are
\begin{eqnarray}
D_{s0}^{*}(2317)\rightarrow \gamma D_{s}^{*} \nonumber \\
X (3700) \rightarrow \gamma J/\psi \nonumber
\end{eqnarray}
and the technical way to evaluate them is considering the loop
diagrams of Fig. \ref{fig1}

\begin{figure}[htb]
\begin{center}
\includegraphics[width=7cm]{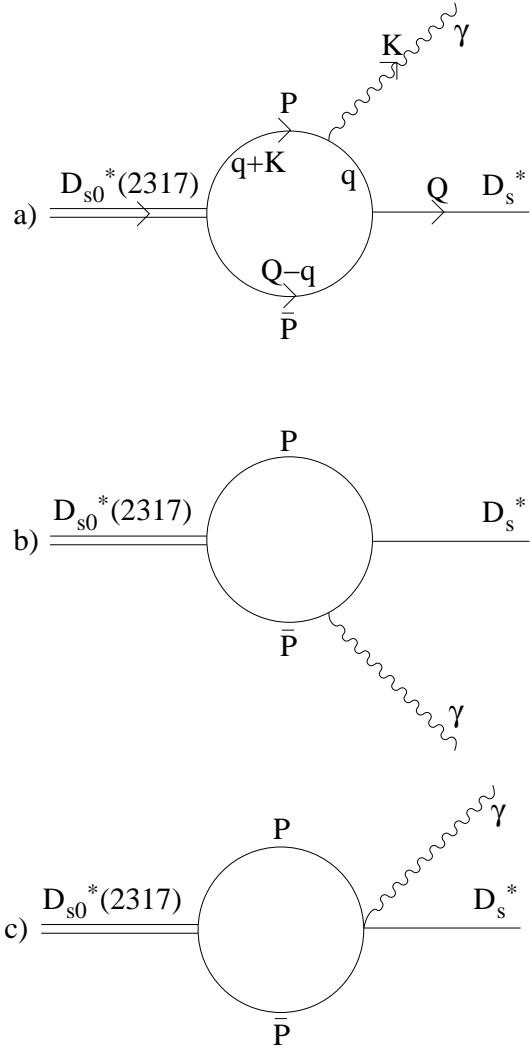}
%\vglue -0.5cm
 \caption{\small Diagrams considered in the
evaluation of the radiative decay $D_{s0}^{*}(2317)\rightarrow
\gamma D_{s}^{*}$. $P$ and $\bar{P} $ are the pseudoscalar mesons that
couple to the $D_{s0}^{*}(2317)$. } \label{fig1}
\end{center}
\end{figure}

These are the same diagrams used in the evaluation of the radiative
decay of the $D_{s0}^{*}(2317)$ as a $DK$ molecule in \cite {prd76}. The
differences are that here we do not need the wave function of the
$D_{s0}^{*}(2317)$, we only need the couplings of the resonance to the
$P\bar{P}$ channels which are obtained in the study of these
resonances as dynamically generated in \cite{daniel}. Another
difference is that we have more channels than $DK$, and this has
some numerical effects in the results because of cancellations of
terms in the most important channels  $D^+ K^0$, $D^0 K^+$. Finally,
although it provides a small contribution in \cite {prd76}, there is
a term where the photon couples to the $D_{s}^{*}$. This term
involves a transition $D_{s0}^{*}(2317)\rightarrow D_{s}^{*}$ which
is allowed for a virtual $D_{s}^{*}$ as it would be the case here. Such
terms mix the longitudinal part of the vector meson propagator with the scalar
meson. However, in order to prevent the appearance of a pole of a
scalar in the vector meson propagator, in a covariant formalism like
the one we use, this transition amplitude must vanish for $P^2=m_S^2$,
with $P$ and $m_S$ the total momentum and the mass of the scalar meson,
respectively \cite{yad,yad2}. This is discussed in a related work on the
radiative decay of axial vector mesons \cite{Roca:2006am}, but we address the problem
in detail in appendix A. There is also a diagram with the photon attached to the
scalar resonance (for charged states), which is shown to vanish in the amplitude in \cite{prd76}
due to the lorentz condition of the vector meson.

The two diagrams discussed above, together with those considered in figure \ref{fig1}, provide
a set of gauge invariant terms, as shown explicitly in \cite{prd76}. The two terms with the
photon coupling to the external particles in the loop diagram play a role in the gauge invariant
test of the theory, as shown in \cite{prd76}, but they vanish in the radiative decay amplitude,
as we show in appendix A.

The procedure to evaluate the radiative decay followed here for the
dynamically generated scalar resonances has been tested with
success in the decays $\phi \rightarrow f_0 (980) \gamma$ and $\phi \rightarrow a_0 (980)\gamma$
\cite {plb470,Oset:npa729,V.E. Markushin} with the $f_0$ and $a_0$
resonances dynamically generated  from the interaction of the
lowest order meson-meson chiral Lagrangian \cite{Oller:1997ti}. The
present reaction is the time reversal reaction, in the charmed sector, of the radiative
$\phi$ decay into a scalar and a photon.
The same ideas presented here are used in the study of the radiative decay of the 
$f_0(980)$ and $a_0(980)$, as dynamically generated resonances, into $\gamma\rho$
and $\gamma\omega$ in \cite{hanhart}.

The channels to which the $D^*_{s0} (2317)$ and the $X(3700)$
resonances have appreciable couplings in \cite{daniel} are the following: 

$D^*_{s0} (2317): D^+K^0,D^0 K^+,D_s^+ \eta,D_s^+ \eta_c$

$X(3700):D^+D^-,D^0\bar{D^0},D_s^+D_s^-$

We shall demonstrate that, using arguments of gauge invariance, we
can overcome the evaluation of the diagram c) of Fig \ref{fig1} and, as a
consequence, we must only evaluate the diagrams of Fig \ref{fig2} for the
$D^*_{s0}(2317)$  and of Fig \ref{fig3} for the $X(3700)$

\begin{figure}[htb]
\begin{center}
\includegraphics[width=9cm]{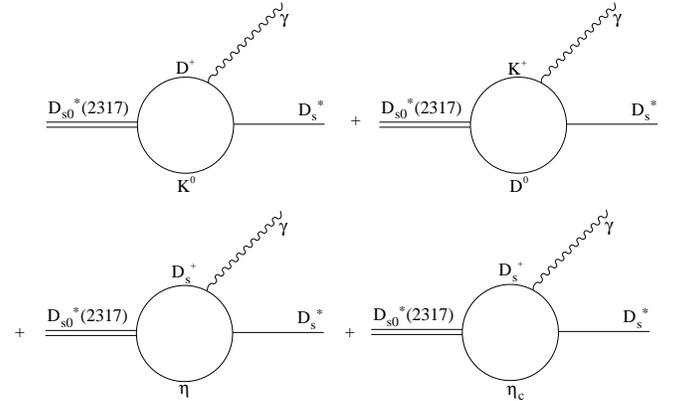}
%\vglue -0.5cm
\caption{\small Diagrams needed in the
evaluation of the $D_{s0}^{*}(2317)$ radiative decay. } \label{fig2}
\end{center}
\end{figure}

\begin{figure}[htb]
\begin{center}
\includegraphics[width=9cm]{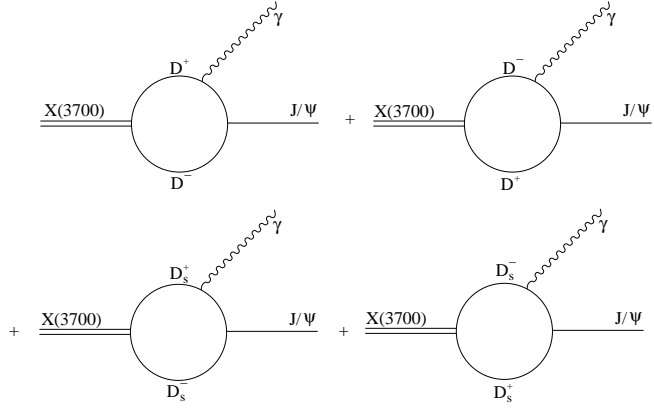}
%\vglue -0.5cm 
\caption{\small Diagrams needed in the
evaluation of the $X(3700)$ radiative decay. }\label{fig3}
\end{center}
\end{figure}

Let us proceed to the explicit evaluation of the diagrams. The
amplitude of the diagram of Fig \ref{fig2} a) is readily evaluated as:

\begin{eqnarray}
-i {\cal T}&=&\int{d^4q\over(2\pi)^4}(-i)g_{D_{s0}^*(2317)\rightarrow P\bar{P}}\nonumber \\
&\times & {i\over(q+K)^2-m_1^2+i\epsilon}{i\over q^2-m_1^2+i\epsilon}\nonumber \\
&\times & {i\over(Q-q)^2-m_2^2+i\epsilon}(-i)eQ_1\epsilon_\nu(\gamma)(q+q+K)^\nu\nonumber \\
&\times & (+i){1\over\sqrt{2}}{M_V G_V \over f_\pi f_D}\epsilon_\mu(D_s^*)(q-Q+q)^\mu \lambda_V \label{ampli}
\end{eqnarray}
where $m_1$ and $m_2$ are the masses of the upper and lower pseudoscalar
mesons in the loop diagram, $eQ_1$ is the charge $(e>0)$ of the
upper pseudoscalar meson, $M_V$ the mass of the vector meson and
$G_V$ the VPP coupling ($G_V=55$ MeV). The constant
$g_{D^*_{s0}(2317)\rightarrow D^+ K^0}$ provides the coupling of the
resonance to the $D^+ K^0$. In Table 1 we show the calculated couplings 
for the poles from \cite{daniel} located at 2317.25 MeV and (3718.93-i0.06) MeV.
For this last one we have neglected its imaginary part.

The coupling of the vector meson to the pseudoscalars is obtained
from the $SU(4)$ generalization (see \cite{daniel}) of the ordinary
$SU(3)$ Lagrangian.

\begin{eqnarray}
 {\cal{L}}_{PPV}&=&\frac{-i g}{\sqrt{2}}Tr([\partial_\mu \phi,\phi]V^\mu)
\end{eqnarray}
with $g=-M_V G_V /f^2$ and the $SU(4)$ matrices $\phi$ and $V^\mu$ given by

\begin{widetext}
\begin{eqnarray}
\phi &=&\left( \begin{array}{cccc}
 {\pi^0 \over \sqd}+{\eta \over \sqs}+{\eta_c \over \sqrt{12}} & \pi^+ & K^+ & \bar D^0 \\ & & & \\
 \pi^- & {-\pi^0 \over \sqd}+{\eta \over \sqs}+{\eta_c \over \sqrt{12}} & K^0 & D^- \\& & & \\
 K^- & \bar K^0 & {-2\eta \over \sqs}+{\eta_c \over \sqrt{12}} & D_s^- \\& & & \\
 D^0 & D^+ & D_s^+ & {-3\eta_c \over \sqrt{12}} \\ \end{array} \right)   \\
V^\mu&=&\left( \begin{array}{cccc}
{\rho_\mu^0 \over \sqd}+{\omega_\mu \over \sqs}+{J/\psi_{ \mu} \over \sqrt{12}} & \rho^+_\mu & K^{*+}_\mu & \bar D^{*0}_\mu \\ & & & \\
 \rho^{*-}_\mu & {-\rho^0_\mu \over \sqd}+{\omega_\mu \over \sqs}+{J/\psi_{\mu} \over \sqrt{12}} & K^{*0}_\mu & D^{*-}_\mu \\& & & \\
  K^{*-}_\mu & \bar K^{*0}_\mu & {-2\omega_\mu \over \sqs}+{J/\psi_{\mu} \over \sqrt{12}} & D_{s\mu}^{*-} \\& & & \\
D^{*0}_\mu & D^{*+}_\mu & D_{s\mu}^{*+} & {-3J/\psi_\mu \over \sqrt{12}} \\ \end{array} \right).   
\end{eqnarray}
\end{widetext}
 
In eq. (\ref{ampli}) we have chosen to substitute $f^2$ in the denominator by $f_\pi f_D$
following the prescription in \cite{daniel} of associating $f_\pi$
(the pion decay constant, $f_\pi=$93 MeV) to the light pseudoscalars
and $f_D=$165 MeV to the charmed pseudoscalars.

 All the other
diagrams are easily obtained by changing the couplings of the
resonance to the channel and the constant $\lambda_V$ to account for
the different $VPP$ vertices. The different values of $\lambda_V$ are given in Table \ref{table2}.

\begin{table}
\begin{center}
\caption{Couplings of the resonances to the pseudoscalars.} \label{table1}
\begin{tabular}{c|c|c}
\hline
Resonance & Channel & $g_{R\rightarrow P\bar{P}}$ [MeV] \\
\hline
\hline
$D_{s0}^*(2317)$&$D^+K^0$&-7358 \\
\cline{2-3}
& $K^+D^0$&-7358 \\
\cline{2-3}
& $D_s^+\eta$&5993 \\
\cline{2-3}
& $D_s^+\eta_c$&1541 \\
\hline
X(3700)&$D^+D^-$&7353 \\
\cline{2-3}
&$D^-D^+$&7353 \\
\cline{2-3}
&$D_s^+D_s^-$&6740 \\
\cline{2-3}
&$D_s^-D_s^+$&6740 \\
\hline
\end{tabular}
\end{center}
\end{table}

\begin{table}
\begin{center}
\caption{Coefficients $\lambda_V$ for the coupling of the vector meson to the pseudoscalars.} \begin{tabular}{c|c|c}
\hline
Vector Meson & Channel & $\lambda_V$ \\
\hline
\hline
$D_s^*$&$D^+K^0$&1 \\
\cline{2-3}
& $K^+D^0$&-1 \\
\cline{2-3}
& $D_s^+\eta$&$-2/\sqrt{6}$ \\
\cline{2-3}
& $D_s^+\eta_c$&$2/\sqrt{3}$ \\
\hline
$J/\psi$&$D^+D^-$&$-2/\sqrt{3}$ \\
\cline{2-3}
&$D^-D^+$&$2/\sqrt{3}$ \\
\cline{2-3}
&$D_s^+D_s^-$&$-2/\sqrt{3}$ \\
\cline{2-3}
&$D_s^-D_s^+$&$2/\sqrt{3}$ \\
\hline
\end{tabular}
\label{table2}
\end{center}
\end{table}

By using the Lorentz condition for the photon and the vector meson,

\begin{eqnarray}
\epsilon_\mu(D^{*+})Q^\mu&=&0 \label{lor1} \\
\epsilon_\nu(\gamma)K^\nu&=&0, \label{lor2}
\end{eqnarray}
the amplitude of eq. (\ref{ampli}) is simplified and we obtain:

\begin{eqnarray}
{\cal T}&=& -ig_{D_{s0}^*(2317)\rightarrow P\bar{P}}eQ_1\lambda_V{4\over\sqrt{2}}{M_V G_V\over f_\pi f_D} \epsilon_\mu(D_s^*)\epsilon_\nu(\gamma) \nonumber \\
&\times & \int{d^4q\over(2\pi)^4}{1\over(q+K)^2-m_1^2+i\epsilon}{1\over q^2-m_1^2+i\epsilon}\nonumber \\
&\times &{1\over(Q-q)^2-m_2^2+i\epsilon}q^\mu q^\nu
\end{eqnarray}

Upon integration of the $q$ variable one has the expression:

\begin{eqnarray}
{\cal T}&=&T^{\mu\nu}\epsilon_\mu(V)\epsilon_\nu(\gamma)
\end{eqnarray}
and Lorentz covariance provides the most general form for $T^{\mu\nu}$ as:

\begin{eqnarray}
T^{\mu\nu}&=&a g^{\mu\nu}+b Q^\mu Q^\nu+c Q^\mu K^\nu+d K^\mu Q^\nu\nonumber \\
&+ & e K^\mu K^\nu \label{deco}
\end{eqnarray}

The Lorentz condition of eqs. (\ref{lor1}) and (\ref{lor2}) removes the contributions of the $b$, $c$ and $e$ terms, such that only the $a$ and $d$ terms contribute. In addition, gauge invariance (which is guaranteed when all the terms in fig. \ref{fig1} are accounted for) $T^{\mu\nu}K_\nu$=0, implies $b=0$ and $a+d Q.K=0$ so,

\begin{eqnarray}
a&=&-d Q.K
\end{eqnarray}
such that only one term is needed in the evaluation. We choose to evaluate the $d$ term because it is finite and only comes from the diagrams of figs. \ref{fig2} and \ref{fig3}. The procedure outlined here has been used before in the evaluation of the $\phi\rightarrow\gamma K^0 \bar{K^0}$ decay \cite{close,oller}.

The amplitude $\cal T$ is now easily written as:

\begin{eqnarray}
{\cal T}&=&-d (Q.K g^{\mu\nu}-K^\mu Q^\nu) \epsilon_\mu(V)\epsilon_\nu(\gamma)
\end{eqnarray}

The evaluation of $d$ is straightforward following the Feynman formalism. We write:

\begin{eqnarray}
{1\over abc}&=&2\int_0^1dx\int_0^x dy {1\over \left( a+(b-a)x+(c-b)y\right)^3}
\end{eqnarray}
with
\begin{eqnarray}
a&=&(Q-q)^2-m_2^2 \\
b&=& q^2-m_1^2 \\
c&=& (q+k)^2-m_1^2
\end{eqnarray}

Upon a transformation $q=q'+Q(1-x)-Ky$ we are left with the integral:

\begin{eqnarray}
& & \int{d^4q'\over (2\pi)^4} {\Big(q'+Q(1-x)-Ky\Big)^\mu  \over (q'^2+s+i\epsilon)^3}\nonumber \\
&\times & \Big(q'+Q(1-x)-Ky\Big)^\nu
\end{eqnarray}
with $s=Q^2x(1-x)+2 Q.K(1-x)y-m_2^2+(m_2^2-m_1^2)x$,
which shows that the contribution to the $d$ term comes from:

\begin{eqnarray}
& &  \int{d^4q'\over (2\pi)^4}{K^\mu Q^\nu (1-x)y \over (q'^2+s+i\epsilon)^3}
\end{eqnarray}
where two powers of $q'$ have disappeared from the integral and hence it is convergent. The $q'$ integral is also readily done following the Feynman formalism:

\begin{eqnarray}
\int{d^4q'\over (2\pi)^4}{1 \over (q'^2+s+i\epsilon)^3}&=&{i\pi^2\over(2\pi)^4}{1\over2}{1\over s+i\epsilon}
\end{eqnarray}
and the $d$ coefficient is readily obtained as:

\begin{eqnarray}
d&=&-g_{D_{s0}^*(2317)\rightarrow P\bar{P}}eQ_1\lambda_V {M_V G_V\over f_\pi f_D} {\sqrt{2}\over 8\pi^2} \nonumber \\
&\times & \int_0^1dx\int_0^x dy{(1-x)y\over s+i\epsilon} \label{dterm}
\end{eqnarray}

As mentioned above, one can see, following the same procedure, that the diagram in fig. \ref{fig1} c) only contributes to the $a g^{\mu\nu}$ term of eq. (\ref{deco}) and thus we do not need to calculate it.

The finiteness of the results is also noted in \cite{prd76} where the wave function is governed by a range parameter $\Lambda$, and the results remain finite in the limit of $\Lambda\rightarrow \infty$.

One must sum coherently the contribution of each term of the diagrams in fig. \ref{fig2} to the $d$ coefficient and then the radiative decay width is given finally by

\begin{eqnarray}
\Gamma&=&{1\over 8\pi} {1\over m_{D_{s0}^*(2317)}^2}|\vec{K}| 2 (K.Q)^2 |d|^2
\end{eqnarray}
where $|\vec{K}|$ is photon three momentum in the rest frame of the $D_{s0}^*(2317)$.

The decay of the X(3700) proceeds identically through the same lines using the appropriate 
couplings and masses in the diagrams of fig. \ref{fig3}

\section{Results}

In Table \ref{table3} we show the results for the $d$ coefficient from each term in fig. \ref{fig2}.

\begin{table}
\begin{center}
\caption{Results} \label{table3}
\begin{tabular}{c|c|c}
\hline
Diagram & $d$ [fm] & $\Gamma$ [KeV] \\
\hline
\hline
$K^+D^0$&  0.01284  & 2.518   \\
$D^+K^0$&  -0.00529 & 0.427   \\
$D_s^+\eta$& -0.00197  & 0.059 \\
$D_s^+\eta_c$& 0.00007 & 0.000 \\
\hline
Total & 0.00565  & 0.488 \\
\end{tabular}
\end{center}
\end{table}

As we can see, the largest contribution comes from the $K^+D^0$ intermediate state. The $D^+K^0$ is smaller than the $K^+D^0$ since it involves two heavy pseudoscalar propagators instead of two light ones. Next and weaker than the others is the contribution of the $D_s^+\eta$ channel, and finally the $D_s^+\eta_c$ channel provides a negligible contribution.

Note that the contribution from the two charge partners in the isospin I=0 $DK$ channel is destructive. Had the $D_{s0}^*(2317)$ been an isospin I=1 resonance, the relative couplings to the two channels would have been opposite, making thus a constructive interference and we would have obtained a width of 4 KeV instead of 0.488 KeV, a factor eight times bigger. Furthermore, because of the destructive interference, the effect of the $D_s^+\eta$ channel, which is quite small by itself, becomes relevant. Indeed, if we neglect the channels with the $D_s^+$ meson, the width obtained is $\Gamma=$0.872 KeV a factor 1.8 times bigger than when one takes them into account. Then, one can see that the consideration of all the coupled channels of the approach is quite relevant, which introduces one novel element with respect to the ordinary molecular picture \cite{prd76} where only the dominant $KD$ channel is taken into account.

The results for the X(3700) radiative decay are shown in Table \ref{table4}. We see that this radiative decay is considerably larger than for the $D_{s0}^*(2317)$. In this case all the terms add constructively.

\begin{table}
\begin{center}
\caption{Results} \label{table4}
\begin{tabular}{c|c|c}
\hline
Diagram & $d$ [fm]  & $\Gamma$ [KeV] \\
\hline
\hline
$D^+D^-$&  -0.00314  & 3.709   \\
$D^-D^+$&  -0.00314 & 3.709  \\
$D_s^+D_s^-$& -0.00129  & 0.622 \\
$D_s^-D_s^+$& -0.00129 & 0.622 \\
\hline
Total & -0.00886  & 29.481 \\
\end{tabular}
\end{center}
\end{table}

Next we perform an analysis of the uncertainties in the results. The fact that we have obtained a very small width, because of strong cancellations, indicates that it should be rather sensitive to uncertainties in the input used for the evaluation.

To evaluate the uncertainties we will follow the same procedure used in \cite{daniel}. We will take a random generated ensemble of sets for the input parameters within a physical allowed range and calculate the radiative decay for each set of parameters in the ensemble. The uncertainties in the results are then given by the standard deviation from the mean value calculated:

\begin{eqnarray}
\sigma^2&=&{\sum_{i=1}^N (\bar \Gamma-\Gamma_i)^2 \over N-1} \label{stddev} \\
\bar{\Gamma}&=&{1\over N}\sum_{i=1}^N \Gamma_i
\end{eqnarray}

Since the radiative decay of the $D_{s0}^*(2317)$ is very small and the uncertainties are of the same order of magnitude, we will separately calculate the standard deviation above and under the mean value. The parameters will be generated within the ranges \cite{daniel}:

\begin{eqnarray}
M_V&=&2060\pm52\textrm{   MeV} \nonumber \\
f_D&=&182\pm36\textrm{   MeV} \nonumber \\
f_\pi&=&100\pm15\textrm{   MeV} \nonumber \\
m_{D_{s0}^*(2317)}&=&2316\pm39\textrm{   MeV} \nonumber \\
g_{D_{s0}^*(2317)\rightarrow DK}&=&-6420\pm1790\textrm{   MeV} \nonumber \\
g_{D_{s0}^*(2317)\rightarrow D_s\eta}&=&5250\pm1430\textrm{   MeV} \nonumber \\
g_{D_{s0}^*(2317)\rightarrow D_s\eta_c}&=&1450\pm470\textrm{   MeV} \nonumber \\
m_{X(3700)}&=&3698\pm35\textrm{   MeV} \nonumber \\
g_{X(3700)\rightarrow D^+D^-}&=&8089\pm3125\textrm{   MeV} \nonumber \\
g_{X(3700)\rightarrow D_s^+D_s^-}&=&5339\pm2100\textrm{   MeV} \nonumber \\
\end{eqnarray}

When we do the exercise for $N$=500 randomly generated parameter sets, we obtain

\begin{eqnarray}
\Gamma_{D_{s0}^*(2317)}&=&0.475_{-0.290}^{+0.831} \textrm{ KeV}
\end{eqnarray}
and

\begin{eqnarray}
\Gamma_{X(3700)}&=&18.45\pm13.00\textrm{  KeV}
\end{eqnarray}

It is instructive to compare our results with those of \cite{prd76} for the $D_{s0}^*(2317)$. They are rather similar. In \cite{prd76} the results vary from 0.47 KeV in some approximations to 1.41 KeV in other approximations. Our uncertainties stem from different sources, couplings, masses, etc, but the range of values obtained is very similar. A comparison of these results with those of different quark models is made in \cite{prd76} and we address the reader to Table III of this reference. The results obtained with the present picture are in general smaller than those obtained in quark models or other pictures. The destructive interference between the two components of the main isospin channel is the main reason for it. Precise experiments on this rate should help us understand better the nature of this resonance. As for the X(3700), its search as a peak in some reactions would be a first step. The search for its radiative decay could follow and, given the large rate predicted, the investigation of this decay channel does not look particularly difficult, specially when the ratio of the radiative decay of the $D_{s0}^*(2317)\rightarrow\gamma D_s^*$ to the $D_{s0}^*(2317)\rightarrow\pi^0 D_s^*$ has already been measured \cite{cleo}.

\section{Conclusions}

We have presented here the evaluation of the radiative decay of the open and hidden-charm scalar mesons which are dynamically generated from the interaction of two pseudoscalar mesons. The calculations have been done for the $D_{s0}^*(2317)$ open-charm scalar state and for the predicted hidden-charm state X(3700), not yet observed. We found very different results for the two states. While the $D_{s0}^*(2317)$ decay into $\gamma D_s^*$ has a width of around 0.5 KeV, the X(3700) has a width into $\gamma J/\psi$ of the order of 20 KeV, a factor forty times bigger. One of the reasons, but not the only one, was the large cancellation between the two charge partners of the isospin component of the $DK$ I=0 state. With the obvious similarities, we also found subtle differences between the $DK$ molecular picture for the $D_{s0}^*(2317)$ state and the dynamically generated picture. The latter one, including more channels than just the $DK$, showed sensitive effects from the $D_s\eta$ state, particularly because of the large cancellation found between the dominant $D^+K^0$ and $K^+D^0$ states. Yet, within the theoretical uncertainties, the final result obtained in the two pictures are rather similar.

We also presented a different technical way to evaluate the amplitudes which makes the formalism simpler and shows immediately the finiteness of the results using arguments of gauge invariance.

Concerning the X(3700) state and its radiative decay, the large width obtained for the decay into $\gamma J/\psi$ should make its observation easy, in principle, and we also recalled that the predictions on this state were rather solid so that it should be observed as a bound state or a strong cusp, in both of which cases the radiative decay could be investigated. The observation of this state with its relatively large radiative width would provide a boost to the idea of the low energy scalar mesons with open-charm and some particular hidden-charm scalar states as dynamically generated resonances.

\acknowledgements{We would like to thank H. Nagahiro and V. E. Lyubovitskij for useful discussions.
One of us, D. Gamermann, wishes to acknowledge support from the ministerio de educacion 
y ciencia in the FPI program. This work was supported in part by DGICYT contract
number FIS2006-03438, the Generalitat Valenciana and the National Natural Science
Foundation of China (No. 10675058)  and Scientific Research
Foundation of Liaoning Education Department (No. 20060490). This research is in part supported by the Japan (JSPS) - Spanish (CSIC) collaboration agreement. This research is  part of the EU 
Integrated Infrastructure Initiative Hadron Physics Project under contract number
RII3-CT-2004-506078.}

\appendix

\section{Photon Coupling to External Lines}

In addition to the diagrams considered in figure \ref{fig1}, one should also include two extra diagrams, for the case
of the decay of charged particles, where the photon couples to the external lines, see figure \ref{fig4}.

\begin{figure}[htb]
\begin{center}
\includegraphics[width=7cm]{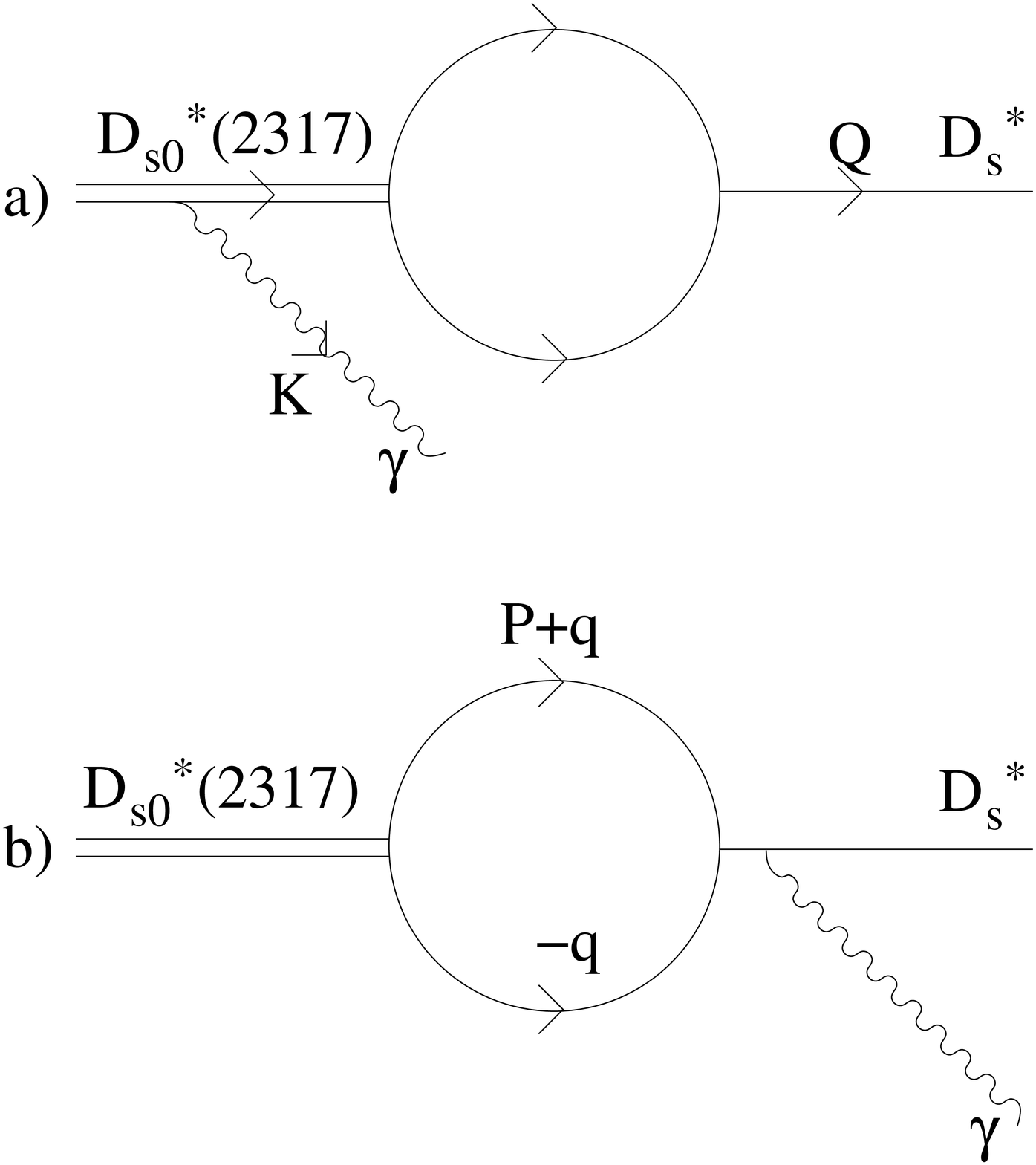}
%\vglue -0.5cm
 \caption{\small Diagrams with the photon conected to the external lines. } \label{fig4}
\end{center}
\end{figure}

These two diagrams are explicitly considered in \cite{prd76} and shown to be relevant in the test of gauge invariance.
However, we show here that they vanish in the amplitude for on-shell $D_{s0}^*(2317)$ and $D_s^*$.

To prove this we first evaluate the loop function in figure \ref{fig4}:

\begin{eqnarray}
J(P^2)P^\mu\epsilon_\mu(D_s^*)=i\int{d^4q\over(2\pi)^4}{g_{D_s^*DK}\over(P+q)^2-m_1^2+i\epsilon}\nonumber\\
\times {g_{D_{s0}^*DK}
\over q^2-m_2^2+i\epsilon}(P+2q)^\mu\epsilon_\mu(D_s^*)
\end{eqnarray}

Both diagrams in figure \ref{fig4} imply vector-scalar mixing, which appears througth the longitudinal
part of the vector meson propagator. Indeed, let us consider the diagrams of figure \ref{fig5}
for the vector meson propagator.

\begin{figure}[htb]
\begin{center}
\includegraphics[width=9cm]{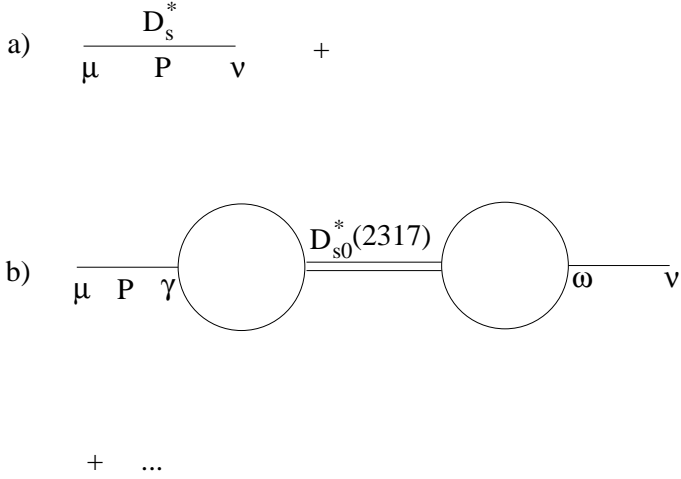}
%\vglue -0.5cm
 \caption{\small Diagrams for the renormalization of the vector meson propagator. } \label{fig5}
\end{center}
\end{figure}

We have:

\begin{eqnarray}
i{\cal D}^{(a)}(P)&=&i\sum_\lambda {\epsilon^\mu(D_s^*)\epsilon^\nu(D_s^*) \over P^2-M^2+i\epsilon} \\
{\cal D}^{(a)}(P)&=&\left( -g^{\mu\nu}+{P^\mu P^\nu \over P^2}\right){1\over P^2-M^2+i\epsilon}\nonumber \\
&+&{P^\mu P^\nu \over P^2M^2}\label{vecprop}
\end{eqnarray}

Where in eq. (\ref{vecprop}) we separated explicitly the propagator into its transverse (first) and longitudinal (second) components. Analogously, figure \ref{fig5}b provides a contribution to the $D_s^*$ propagator given by:

\begin{eqnarray}
i{\cal D}^{(b)}(P)&=&i\sum_\lambda {\epsilon^\mu(D_s^*)\epsilon^\gamma(D_s^*) \over P^2-M^2+i\epsilon}(-i)J(P^2)P_\gamma
i\tilde{\cal D}_{D_{s0}^*(2317)}(P) \nonumber \\
&\times& (-i)J(P^2)P_\omega i\sum_{\lambda `} {\epsilon^\omega(D_s^*)\epsilon^\nu(D_s^*) \over P^2-M^2+i\epsilon} \label{twoloop}
\end{eqnarray}
where $\tilde{\cal D}_{D_{s0}^*(2317)}(P)$ is the propagator of the scalar particle.

One can see that the presence of $P_\gamma P_\omega$ in eq. (\ref{twoloop}) eliminates the contribution
of the transverse part of the vector meson propagator, hence, only the longitudinal part contributes, and we obtain:

\begin{eqnarray}
{\cal D}^{(b)}(P)&=&{P^\mu P^\nu \over M^4}J(P^2)^2{1\over P^2-m_{D_{s0}^*(2317)}^2}
\end{eqnarray}

The iteration of the last diagram of figure \ref{fig5} and the sum of all these terms leads to a geometrical series
which renormalizes the longitudinal part of the vector propagator and leads to:

\begin{eqnarray}
{P^\mu P^\nu \over P^2M^2}&\rightarrow&{P^\mu P^\nu \over P^2M^2}\left({1\over 1-{P^2\over M^2}J(P^2)^2{1\over P^2-m_{D_{s0}^*(2317)}^2}}\right) \nonumber \\
&=&{P^\mu P^\nu \over P^2M^2}{P^2-m_{D_{s0}^*(2317)}^2\over P^2-m_{D_{s0}^*(2317)}^2-{P^2\over M^2}J(P^2)^2}
\end{eqnarray}

Now comes an important renormalization condition which is the physical requirement that the longitudinal part of
the vector meson propagator does not contain a pole of the scalar meson \cite{kaloshi,lope}. This condition is only fulfilled if

\begin{eqnarray}
J(P^2=m_{D_{s0}^*(2317)}^2)&=&0.
\end{eqnarray}

Next we evaluate the two terms in the amplitude of figure \ref{fig4}.

\begin{eqnarray}
-i{\cal T}^{(a)}&=&-ie(2Q+K)^\mu\epsilon_\mu(\gamma)i\tilde{\cal D}_{D_{s0}^*(2317)}(Q)\nonumber \\
&\times& (-i)J(Q^2)Q^\nu\epsilon_\nu(D_s^*)
\end{eqnarray}

This term is zero because of the lorentz condition on the vector meson, $Q^\nu\epsilon_\nu(D_s^*)=0$. This was already realized and used in \cite{prd76}.

Next we look at the diagram which contributes to the amplitude in figure \ref{fig4}b:

\begin{eqnarray}
-i{\cal T}^{(b)}&=&-iJ(P^2)P^\mu\epsilon_\mu(D_s^*)...
\end{eqnarray}

As we can see, independently of the $\gamma VV$ coupling, the term ${\cal T}^{(b)}$ is proportional to $J(P^2=m_{D_{s0}^*(2317)}^2)$, which we have shown before to be zero due to the renormalization condition of
the longitudinal part of the vector meson propagator.

Our procedure to evaluate the amplitude, hence, relies upon:

1) The whole set of diagrams is gauge invariant.

2) The diagrams of figure \ref{fig4} do not give contribution to the amplitude.

3) Only the set of diagrams of figure \ref{fig1} give contribution.

4) Using gauge invariance and the procedure followed through eqs. (9-11), only the $d$ term has to be evaluated, to which 
only the diagrams of figure \ref{fig1}a and \ref{fig1}b contribute.

\end{document}